# Is 1I/2017 U1 really of interstellar origin ?

*Jean Schneider*

*Paris Observatory – Luth UMR 8201*

The detection of the object 1I/2017 U1 (MPEC 2017) on a hyperbolic trajectory presents an interesting question : is it really of interstellar origin or does it genuinly come from the Solar System, accelerated by some planetary encounter ? All subsequent papers (de la Fuente Marcos, Gaidos, Laughlin and Batygin, Mamajek) present it as a probable interstellar object. The issue is important in both cases, because if it were a Solar System object, it would be the first asteroid (*i.e.* non cometary object) from the far outer Solar System. And, in addition, even in that case, its present hyperbolic velocity makes it potentially an interstellar probe if some Lyra-like spacraft (Hein et al. 2017) can chase it and deposit detectors on it (for communication see the talk by D. Messerschmitt at https://tviw.us/2017-presentation-video-archive/ ). Such a probe could, in passing, investigate the Oort cloud *in situ*. If it it of interstellar orgin it provides a new channel of information about planetary systems (Trilling et al. 2017).

Here we investigate the possibility that the asteroid actually is a Solar System object, currently expelled from the Solar System by the recent encouter with a Solar System planet. As pointed out by Laughlin and Batygin (2017), two known planets are sufficiently massive and far from the Sun to expell a Solar System object, namely Jupiter and Neptune. Indeed according to them the factor $f$ given by $f = ((M_{pl} a)/(M_* R_{pl}))^{1/2}$ is significantly larger than 1 to allow an ejection from the Solar System. But this possibility is excluded by the high inclination of the 1I/2017 U1 trajectory (33 deg. de la Fuente Marcos & de la Fuente Marcos 2017 [1]) with respect to the plane of the ecliptic. The asteroid 1I/2017 U1 could not have met these two planets.

But the situation is different with hypothetical Planet Nine (Batygin & Brown 2016) since it is inclined by 18 to 48 deg. with respect to the ecliptic (Malhotra 2017) its $f$ factor is 10 for a mass of $10 M_\oplus$ and a present distance of 500 AU.

Now, one may argue that is it unlekely that the first object accelerated at a significantly hyperbolic velocity is an asteroid and not a comet since up to now no outer Solar Solar system asteroid has never been detected, contrary to comets which are very common. We face here a very low probability configuration, but it is not logically impossible that we may have just detected by chance an object from a very rare category.

The main argument against a Solar System origin of 1I/2017 U1 comes from the location of Planet Nine compared to the incoming direction of the asteroid. In absence of direct detection, this location is not well known, but a well defined region of the sky can be infered from the analysis of the perturbations of Planet Nine to the distribution of Solar System minor planets. Assuming an incoming direction RA,

---

[1] De la Fuente Marcos give an inclination 123°, but that number larger than 90° is to mean thant tht it moves in a direction opposite to Solar System planets. The geometric inclination is 123-90 = 33 °.

DEC = 280 deg, +34 deg. for 1I/2017 U1, one can see from figure 4 of Malhotra (2017) that in any circumstance its past trajectory is at least 35 deg. from the direction of Planet Nine. For a distance of 500 UA of Planet Nine, this means that 1I/2017 U1 passed at least at 450 AU away from Planet Nine, excluding all possibility of an influence of the latter on the asteroid.

In conclusion, unless one invokes the existence of another, yet unknown, planet on the path of the asteroid assumed to come from the Solar System, one is forced to conclude that the latter came directly from the interstellar space.